
\input phyzzx.tex
\date{July 16, 1993}
\voffset=3pc
\hoffset=0.45in

\def\PL  #1 #2 #3 {{\sl Phys.~Lett.}~{\bf#1} (#3) #2 }
\def\NP  #1 #2 #3 {{\sl Nucl.~Phys.}~{\bf#1} (#3) #2 }
\def\PR  #1 #2 #3 {{\sl Phys.~Rev.}~{\bf#1} (#3) #2 }
\def\PRD #1 #2 #3 {{\sl Phys.~Rev.~D} {\bf#1} (#3) #2 }
\def\PRB #1 #2 #3 {{\sl Phys.~Rev.~B} {\bf#1} (#3) #2 }
\def\PP  #1 #2 #3 {{\sl Phys.~Rep.}~{\bf#1} (#3) #2 }
\def\MPL #1 #2 #3 {{\sl Mod.~Phys.~Lett.}~{\bf#1} (#3) #2 }
\def\CMP #1 #2 #3 {{\sl Comm.~Math.~Phys.}~{\bf#1} (#3) #2 }
\def\PRL #1 #2 #3 {{\sl Phys.~Rev.~Lett.}~{\bf#1} (#3) #2 }
\def\TMP  #1 #2 #3 {{\sl Theor.~Math.~Phys.}~{\bf#1} (#3) #2 }
\def\JMP  #1 #2 #3 {{\sl Jour.~Math.~Phys.}~{\bf#1} (#3) #2 }
\def\IJ  #1 #2 #3 {{\sl Int.~Jou.~Mod.~Phys.}~{\bf#1} (#3) #2 }

\REF\wilson{K.~Wilson, \PRB 4 3185 1971  .}
\REF\rec{K. Wilson
and J. Kogut, \PP 12 75 1974 ;
P. ~Bleher and Y. ~Sinai, \CMP 45 247 1975 ; P.~Collet and
J. P. ~Eckmann, \CMP 55 67 1977 ; K. Gawedzki
and A. Kupiainen, Les Houches 1985, K. Osterwalder and R. Stora, Editors }
\REF\dyson{F. Dyson, \CMP 12 91 1969 ; G. Baker, \PRB 5 2622 1972 . }
\REF\preprint{Y. Meurice and V. Rodgers Univ. of Iowa Preprint in preparation.
}
\REF\taibleson{M. Taibleson, {\it Fourier Analysis on Local Fields}, Princeton
University Press.}
\REF\walsh {For an introduction to Walsh functions see N. Fine,
{\sl Trans. Amer
. Math.}, {\bf 65 } (1949) 372.}
\REF\bleher{P.Bleher, \CMP 84 557 1982 ; E. ~Lerner and M. ~Missarov, \TMP 78
177 1989 . }
\REF\meurice{Y.~Meurice, \PL 265B 377 1991 ; J.L.~Lucio and Y. ~Meurice,
\MPL 6 1199 1991 . }
\REF\serre{J.P. Serre, {\it A Course in Arithmetics}, (Springer, New York,
1973). }
\REF\progress{Y. Meurice and G. Ordaz, work in progress.}
\REF\conje{Y. Meurice, \MPL 7 3331 1992  .}

\Pubnum={U. of Iowa 12-93 }
\titlepage
\title{ A Perturbative Improvement of
the Hierarchical Approximation}
\author{Yannick Meurice}
\address{Department of Physics and Astronomy, University of Iowa,
Iowa City, Iowa 52242, USA}
\vfil
\abstract
We propose a perturbative improvement of the hierarchical approximation
for gaussian models. The procedure is based on a relabeling of the momenta
which allows one to
express the symmetries of the hierarchical model using a simple
multiplication group.
The representations of this group are used
to expand the action. The perturbative expansion is treated as a problem
of symmetry breaking using Ward identity techniques.
\endpage

\chapter{Introduction}

In many respects,
gauge theories provide
a satisfactory framework to describe the interactions among elementary
particles at energies accessible with existing colliders.
Nevertheless, the perturbative methods which are spectacularly accurate
for QED at low energy are not adequate to take into account
large distance effects
among strongly
interacting particles.
The renormalization group (RG) method is an essential tool to
understand and describe these phenomena and, more generally,
the infra-red behavior of a
field theory.
However, the practical implementation of this method requires approximations
which are usually
difficult to control or improve. The goal of this paper is to discuss the
improvement
of such an approximation in a very simple scalar theory.

In his remarkable analysis of the partition function of the scalar
field theory, K. Wilson \refmark\wilson
has shown
that by replacing some of the contributions by order of magnitude
estimations, it
was possible to obtain a very simple renormalization group transformation
called the approximate
recursion formula. This recursion formula played an important role in the
development of the
RG method
because the basic ideas (fixed points, relevant directions..) were
not too difficult to work out in detail\refmark{\wilson , \rec }
in this simplified RG transformation. Similar recursion formulas hold exactly
for hierarchical models.\refmark\dyson
For this reason, we call the approximation made by Wilson the hierarchical
approximation.

An interesting feature\refmark\dyson of the hierarchical models is their large
group of symmetry.
An appropriate
use of these symmetries allows, in the case of Ising spins,
to cut down the time
necessary to
calculate numerically\refmark\preprint
the partition function with $2^n$ sites,
from $T=2^{2^n}$ to $(log(T))^2$,
without making
any approximations. In our attempt to improve the hierarchical
approximation, we would like to keep
the largest possible subgroup of unbroken symmetries at each stage of
the calculation.

The improvement of the hierarchical approximation is in principle
straightforward:
we have to restore
the details erased in Wilson's discussion.\refmark\wilson However there are
several practical
problems which
will be encountered. First, there is a bookkeeping problem: we need to
write $\it all$ the corrections.
Second, the is a priority problem: we would like to know which corrections
are the most important
in order to calculate them first. Finally, there is the feasibility
of the calculation itself.

We present here a method which overcomes these problems in a natural way. For
simplicity,
we have restricted
the discussion to a gaussian model on a finite one-dimensional
lattice with $2^n$ sites.
The Fourier modes
of the scalar field are denoted
$\Phi _k $ where $\Phi _k^{\ast} =\Phi _{-k}$ and $k$ are integers
used to express the momenta
in ${2\pi \over {2^n}}$ units.
These integers are understood modulo $2^n$ in the following (periodicity
in momentum space).
The action reads
$$ S={1\over 2^{n+1}} \sum_{k=1}^{2^n} g(k) \Phi_k \Phi_{-k} \eqno(1)$$
where $g(k)$ is even, real and positive .

We proceed in three steps. First, we relabel the momenta in a way which is
convenient to read their
``shell'' assignment in Wilson's discussion (section 2). This
relabeling allows us to introduce
a group of transformation whose orbits are precisely these shells.
The representations
of this group are known\refmark\taibleson
 and can be used to expand the kinetic term , i.e.,
the function $g(k)$ in Eq. (1), for
each of the shells.
This solves the bookkeeping problem. Nicely enough, the
classification of the representations
of the group mentioned above comes with an index indicating its resolution
power (called degree
of ramification). This naturally provides the successive orders of our
perturbative expansion.
We then show that if we only retain the trivial representation in the
expansion, we obtain
the hierarchical approximation (section 3). In this limit, the group
of transformation is a
symmetry of the action which can be identified with the symmetry group of the
hierarchical model mentioned above.
The improvement of the the hierarchical approximation can thus be
treated as a symmetry breaking
problem. We describe the first corrections using the familiar apparatus
of the Ward identities
in section 4.

We emphasize that this procedure takes advantage of the symmetry present at
order zero in an optimal
way. We have checked in simple examples that
the largest corrections correspond to the smallest breaking of the
symmetry.
The extension of this construction for $D$-dimensional models is rather
straightforward
and mentioned at the end
of the paper. The use of this scheme for interactive theories
(Landau-Ginzburg or Ising type)
is presently under study and briefly discussed at the end of section 3.

\chapter{A Convenient Relabeling of the Momenta}

In this section, we introduce a relabeling of the momenta $k$ appearing
in Eq. (1). For this purpose, we use a set orthonormal
functions which is a discrete version of the Walsh system\refmark\walsh
 motivated
by Wilson shell decomposition (see section II of Ref.[1] for detail).
We first define
$$ \Psi _0 (k) \equiv \cases{  1\ {\rm{if}} \ k\ =\ -2^{n-2} +1,....,2^{n-2} -1
\cr
                               \omega \ {\rm{if}}\ k\ =\ 2^{n-2} \cr
                              \omega ^{\ast} \ {\rm if} \ k \ = \ -2^{n-2} \cr
                               0 \ {\rm otherwise} } \eqno(2)  $$
and
$$ \Psi _1 (k)\equiv 1- \Psi _0 (k)\  \eqno(3)$$
with the notation $\omega \equiv {{1+i}\over 2}$.
\endpage

\noindent
For a given integer
$a=a_0 + a_1 2^1 + a_2 2^2+........+a_{n-1} 2^{n-1}$ with $a_l =\ 0\ {\rm or }\
 1$
we define
$$f_a(k)\equiv \prod_{l=0}^{n-1} \Psi _{a_l} (2^l k)\ . \eqno(4)$$
It is clear that $f_a^{\ast}(k) =f_a(-k)$ and we can check that
$$\sum_k f_a(k) f_b ^{\ast}(k) = \delta _{a,b}\ .\eqno(5)$$
A more detailed analysis shows that $f_a (k)$ is non-zero only when
$k=\pm k[a]$ for a function $k[a]$ which will
be specified. More precisely, it is possible to write
$$f_a(k)= \omega \delta _{k,k[a]} + \omega ^{ \ast} \delta_{k,-k[a]}\
.\eqno(6)$$
This relation {\it defines} a one-to-one map $k[a]$.
In the following, $a$ will also be treated as an integer modulo $2^n$.
We can now expand
$$\Phi _k = \sum_{a=1}^{2^n} c_a f_a (k)\ .\eqno(7)$$
With this and Eqs. (5) and (6) we can rewrite
$$S={1\over 2^{n+1}} \sum_{a=1}^{2^n} \tilde g(a) c_a ^2 \eqno(8)$$
where $\tilde g (a) \equiv  g(k[a])$.
By construction, the $c_a$ are real field variables. For convenience
we shall also use their complex form
$$\sigma _a \equiv {1\over 2}(c_a +c_{-a}) + {i \over 2} (c_a - c_{-a})
\ .\eqno(9)$$

We can now explain the correspondence between this relabeling and Wilson's cell
decomposition. Clearly, if $a_0=1$, $f_a (k)$ is supported
in the high momentum
region. More precisely, the 0-th shell, i.e, the one integrated first
in the RG procedure, consists in configurations which can be expanded
in terms of the $f_{1+a_1 2+...} (k)$.
Similarly, the modes corresponding tho the $l$-th shell are made out of the
the $f_{ 2^l + a_{l+1} 2^{l+1} +....} (k)$.

\chapter{The Hierarchical Approximation and its Systematic Improvement}

In the previous section, we have introduced new field variables
$c_a$ corresponding to the $l$-th shell
when $a$ can be divided by $2^l$ but not by $2^{l+1}$.
This property is not affected if $a$ is multiplied (modulo $2^n$)
by any odd number. Note also that the odd numbers form an abelian group
with respect to the multiplication modulo $2^n$.
The orbit of this group within the integers modulo $2^n$ are precisely
the sets of numbers that we have put in correspondence with the shells.
The representations of this group have been studied and classified.
The results can be found in a book by Taibleson.\refmark\taibleson

In order to use Taibleson results, we have
to embed the labels introduced above and denoted $a$, in the
$2$-adic integers. Such a technique\refmark\bleher
has already been used, for instance, to
discuss the random walk representation of the
hierarchical model.\refmark\meurice
When $a$ can be divided by $2^l$ but not by $2^{l+1}$, we say that the
2-adic norm, noted $|a|_2$, is $2^{-l}$. In the infinite volume limit,
or in other words when $n$ tends to infinity, the multiplicative group
of the odd numbers is called the 2-adic units. The representations of
this group will be denoted $\Pi _s$. This means that
if $u_1$ and $u_2$ are 2-adic units,
then $\Pi _s (u_1 u_2) = \Pi _s (u_1) \Pi _s (u_2)$.
The label $s$
specifies the representation in a way which will be explained below.

It is easy to construct explicitly the representations $\Pi _s$.
A 2-adic
unit can be written\refmark\serre
as $u=\pm Exp(4z)$ where $z$ is a (2-adic) integer and $Exp$ the 2-adic
exponential.
$\Pi _s (u)$ is even or odd under multiplication by $- 1$.
On the other hand, $z$ is an additive parametrization and
the $z$ dependence of $\Pi _s$ will be of the
form $e^{i{2\pi z  q\over {2^r}}}$ where $q$ is an odd
integer and $r$ a positive integer. Taibleson calls $r+2$
the degree of ramification.
In summary, the label $s$ is a short notation for the parity,
$r$ and $q$ an odd integer modulo $2^r$.

We can now use these representations to expand the the kinetic
term function $\tilde g (a)$ in each shell. For a given shell $l$,
the $a$ have the form $2^l u$ (so $|a|_2=2^{-l}$) and we can write,
$$ \tilde g (2^l u) = \sum _s g{\ \atop {l,s}}  \Pi _s (u) \eqno(10) $$
At finite volume, i.e., at finite $n$, the units are understood
modulo $2^{n-l}$ and consequently
the sum over the representations
$s$ is restricted to $r \leq
n-l-2$. The numerical coefficients $g {\ \atop {l,s}}$ are easily calculable
using the orthogonality relations among the representation.

The hierarchical approximation is obtained by retaining only the trivial
representation in the expansion (10). In this approximation,
and using the definition introduced in Eq.(9), the action reads
$$S={1\over 2^{n+1}} \sum_{l=0}^{n-1} g{\ \atop {l,+,0}} \sum_{a:|a|_2 =
2^{-l}}
\sigma _a \sigma _{-a} \eqno(11)$$
After a Fourier transform, we obtain a hierarchical model having the
general form written by Dyson in Ref.[3], as discussed below.

The classification of the representations of the 2-adic units suggest
that we improve the hierarchical approximation by taking into
account the additional terms in the expansion (10) ${\it
 order\ by\ order\ in\ the\ degree\ of\ ramification}$. Intuitively, this
corresponds to the fact that the degree of ramification measures
the ``power of resolution'' of the representation. Numerically, this
works quite well. For instance, for $g(k)=1-cos({k 2\pi \over {2^n}})$,
we obtain with $n=4$ and $l=0$,
$g{\ \atop {+,0}}=1.63,\ g{\ \atop {-,0}}=0.31,\ g{\ \atop{+,1}}=0.25$ and
$g{\ \atop{-,1}}=-0.05$.
The subscripts denote the parity and $r$ respectively.
Obvious indices have been skipped.

The actions (8) and (11) take a more familiar form after Fourier transformation
in $a$.
In general, the $l$-th momentum shell is responsible for
interactions among the averages of the Fourier transformed fields
inside boxes of size $2^l$. In the hierarchical approximation, the interactions
depend
only on the position of these boxes inside boxes of size $2^{l+1}$. When the
corrections are
introduced up to $r=r_{max}$, the interactions depend on the position inside
boxes of
size $2^{l+2+r_{max}}$. If local interactions in these new fields are
introduced, then
it is easy to generalize the numerical\refmark\preprint or
diagrammatic\refmark\progress
treatment for small $r_{max}$.

\chapter{The Improvement of the Hierarchical Approximation As a Symmetry
Breaking Problem}

It is important to realize that the hierarchical approximation of $S$ given in
Eq.(11) is invariant under the transformation
$$\sigma_a \longrightarrow \sigma_{ua} \eqno(12)$$
for any odd number $u$.
When the other terms of the expansion are incorporated, this symmetry
is broken by each term in a definite way.
This allows us to use Ward identities techniques.
We discuss the simplest case below.

Suppose we want to calculate the 2-point function, using the perturbative
expansion described
in the previous section.
First we use the new variables $c_a$ and the inverse of the map $k[a]$
defined in section 2
to write
$$\VEV{\Phi _k \Phi_{-k} } = \VEV{ c^2 _{a[k]} } .\eqno(13)$$
In the hierarchical approximation, the value of this expression depends
only on the momentum shell specified by $|a|_2$. In other words,
$$\VEV{ c^2 _{ua}}  _0 = \VEV{c^2 _{a}} _0  \eqno(14)$$
where $\VEV{\ } _0$ means that the quantity is evaluated, at order zero,
or in other words, in the
hierarchical approximation.

Suppose that we now include a correction $\delta \tilde g(a) $ to
the approximation of $\tilde g(a)$.
Then in first order in this perturbation, we
recover the a momentum dependence within the shells given by
$$ \VEV{ c^2 _{ua} } _1 = \VEV{c^2 _{a}} _1-{1 \over 2^{n+1}}
\sum _b (\delta \tilde
g(ub)- \delta \tilde g(b) ) \VEV{c_b^2 c_a^2} _0 \eqno(15) $$
In the model considered here, the  $\VEV{c_b^2 c_a^2} _0$ contribution
can be evaluated straightforwardly and we can check that we recover
the first term in the expansion of $(1/\tilde g(ua))-(1/\tilde g(a)))$.
The important point is that the corrections are evaluated using the
unperturbed action.
It is clear that similar methods can be used for higher point functions
and in the interacting case.

\chapter{Conclusions and Perspectives}

We have proposed a systematic improvement of the hierarchical
approximation for the gaussian model in one dimension. This
procedure exploits in an optimal way the symmetries of the
approximation and make sense as a perturbative expansion.

An extension to $D$-dimension can be constructed easily by noticing
the approximate correspondence between the integration over successive shells
and the block-spin method. On a $D$-dimensional cubic lattice,
we can decompose the block-spin procedure into $D$ steps (one in each
directions). This suggests the introduction of  functions
$$\Psi _0 ^j (k_1,...k_D)\ \equiv \Psi _0 (k_j) \eqno(16) $$
and the replacement of the definition (4) by
$$f_a(k_1......k_D)=\Psi_{a_0}^1 (k_1).....\Psi_{a_{D-1}}^D(k_D)
\Psi_{a_D}^1 (2k_1).....\Psi_{a_{2D-1}}^D(2k_D).....\eqno(17)$$
This procedure keeps the computational advantages of the $D=1$ case.

Our general objective is the construction of a perturbative expansion
for which the RG method and the $\epsilon $-expansion
are very well understood in the 0-th order approximation.
A first test of the reliability of the method will be the calculation
of the first corrections to the critical exponents. In the longer term,
we also expect the method to shed some light on Polyakov's conjecture
for the 3$D$ Ising model.\refmark\conje

\ack
This work was completed during my stay at the Aspen Center for Physics
in June and July 1993. I would like to thank the organizers and participants
for the stimulating atmosphere provided there.

\refout
\end